\documentclass[showpacs,preprintnumbers,amsmath,amssymb,APSl,prd,nofootinbib,superscriptaddress, twocolumn]{revtex4-1}

\usepackage{dcolumn}% Align table columns on decimal point
\usepackage{bm}% bold math
\usepackage{ifpdf}
\usepackage{hyperref}
\usepackage{bm}
\usepackage{xcolor,color,graphicx,graphics}
\usepackage[spanish,english]{babel}%, portuguese
\usepackage[latin1]{inputenc}
\usepackage[OT1]{fontenc}
\usepackage{latexsym,amssymb,amsmath,amsfonts}
\usepackage{makeidx}
\usepackage{epsfig,subfigure}
\usepackage{natbib}
\usepackage{epstopdf}
\usepackage{mathrsfs}
\hypersetup{colorlinks=true, linkcolor=blue, citecolor=green}
\usepackage{enumerate}
\usepackage{xcolor}
\usepackage{multirow}
 \usepackage{array}
\definecolor{red}{rgb}{1,0,0}

\def\p{\partial}

\def\+{^\dagger}

\def\<{\leftarrow}
\def\>{\rightarrow}

\def\({\left(}
\def\){\right)}

\def\a{\alpha} \def\b{\beta}   \def\e{\ephilon}
\def\m{\mu} \def\n{\nu} \def\r{\rho}   
\def\k{\kappa}\def\t{\tau}

\newcommand{\bi}{\begin{itemize}} 				\newcommand{\ei}{\end{itemize}}
\newcommand{\benu}{\begin{enumerate}} 		\newcommand{\enu}{\end{enumerate}}
\newcommand{\bd}{\begin{dinglist}{0}}     \newcommand{\ed}{\end{dinglist}}
\newcommand{\bfig}{\begin{figure}[htbp]}  \newcommand{\efig}{\end{figure}}
        			
\newcommand{\bc}{\begin{center}} 				  \newcommand{\ec}{\end{center}}
\newcommand{\be}{\begin{equation}} 				\newcommand{\ee}{\end{equation}}
\newcommand{\bsub}{\begin{subequations}}  \newcommand{\esub}{\end{subequations}}
\newcommand{\ben}{\begin{eqnarray}} 			\newcommand{\een}{\end{eqnarray}}
\newcommand{\ba}[1]{\begin{array}{#1}} 		\newcommand{\ea}{\end{array}}
\newcommand{\bea}{\begin{equation}\begin{array}{rcl}}
\newcommand{\eea}{\end{array}\end{equation}}

\def\k{\kappa}\def\r{\rho}\def\t{\theta} \def\x{\xi} \def\p{\pi} \def\m{\mu} \def \e{\epsilon} \def\a{\alpha} \def\n{\nu}

\begin{document}

\title{Pre-main sequence evolution of low-mass stars in Eddington-inspired Born-Infeld gravity}

\author{Merce Guerrero} \email{merguerr@ucm.es}
\affiliation{Departamento de F\'isica Te\'orica and IPARCOS,
	Universidad Complutense de Madrid, E-28040 Madrid, Spain}	
	
\author{Diego Rubiera-Garcia} \email{drubiera@ucm.es}
\affiliation{Departamento de F\'isica Te\'orica and IPARCOS,
	Universidad Complutense de Madrid, E-28040 Madrid, Spain}

\author{Aneta Wojnar}
\email{aneta.magdalena.wojnar@ut.ee}
\affiliation{Laboratory of Theoretical Physics, Institute of Physics, University of Tartu,
W. Ostwaldi 1, 50411 Tartu, Estonia
}

\begin{abstract}
    We study three aspects of the early-evolutionary phases in low-mass stars within Eddington-inspired Born-Infeld (EiBI) gravity, a viable extension of General Relativity. These aspects are concerned with the Hayashi tracks ({\it i.e.} the effective temperature-luminosity relation); the minimum mass required to belong to the main sequence; and the maximum mass allowed for a fully convective star within the main sequence. In all cases we find a dependence of these quantities not only on the theory's parameter, but also on the star's central density, a feature previously found in Palatini $f(R)$ gravity. Using this, we investigate the evolution of these quantities with the (sign of the) EiBI parameter, finding a shift in the Hayashi tracks in opposite directions in the positive/negative branches of it, and an increase (decrease) for positive (negative) parameter in the two masses above. We use these results to ellaborate on the chances to seek for traces of new physics in low-mass stars  within this theory.

\end{abstract}

\maketitle

\section{Introduction}

Despite the many tests that Einstein's General Theory of Relativity (GR) has successfully passed \cite{Will:2014kxa}, over the last decades a plethora of modified theories of gravity has been proposed in order to address its shortcomings \cite{Capozziello:2011et,Nojiri:2017ncd}. These include the yet undetected dark energy/matter sources needed for the consistence of the cosmological concordance model \cite{Bull:2015stt}, or the existence of space-time singularities at high-energy scales \cite{Senovilla:2014gza}, such as the one at the center of black holes and the Big Bang singularity.  Attempts to modify GR must come along with the necessity to comply with its well tested weak-field limit, while deviations with respect to its predictions must be searched in those domains where the strength of the gravitational interaction grows large enough, for instance, via gravitational waves out of binary mergers \cite{Ezquiaga:2020dao}, gravitational lensing and shadows \cite{EventHorizonTelescope:2020qrl}, or in the structure of neutron stars \cite{Olmo:2019flu}.

From an astrophysical point of view, neutron stars are suitable objects to test modified gravity via the opportunity in the combination of electromagnetic radiation and gravitational waves that the newly born field of multimessenger astronomy offers \cite{LIGOScientific:2017vwq}. Among the open problems in this field, one can mention the degeneracy of the mass-radius relations due to the fact that the equation of state at supranuclear densities is unknown \cite{Camenzind}, the theoretical difficulties to meet the observations of neutron stars above two solar masses \cite{Linares:2018ppq,Antoniadis:2013pzd}, or the so-called mass gap problem, namely, the existence of objects above the neutron star mass limit but below the lightest black hole mass, as manifested in the observation of gravitational waves from the merging of two objects with 2.6 and 23 $ M_\odot $, respectively \cite{LIGOScientific:2020zkf}.

We consider here the alternative path of focusing on other astrophysical objects for which their internal structure and equation of state are better known. Even though in such objects the gravitational interaction is less strong as compared with neutron stars and black holes, modified gravity effects are able to induce extra terms to the Poisson equation, see e.g. \cite{Sakstein:2018fwz}. This leads to a different stellar structure whose associated macroscopic features can be tested. As examples, we highlight the predictions for the time scales and effective luminosities of both main sequence stars \cite{Chowdhury:2020wfr} and  sub-stellar objects such as as brown dwarfs \cite{Benito:2021ywe} and giant gaseous planets \cite{Wojnar:2021xbr}, lithium abundance and age determination for white dwarfs and low mass stars (LMS) 
\cite{Wojnar:2020frr,Wojnar:2020ckw,lupa}, the minimum hydrogen burning mass for high-mass brown dwarfs to belong to the main sequence, \cite{Sakstein:2015zoa, Sakstein:2015aac, Crisostomi:2019yfo, Olmo:2019qsj,Rosyadi:2019hdb}, early evolutionary tracks of LMS \cite{Wojnar:2020txr}, and even tests with exoplanets \cite{olek1,olek2,olek3} are at our disposal.

The main aim of the present work is to highlight the most important phases of the early evolution of LMS,  within a suitable extension of GR dubbed as Eddington-inspired Born-Infeld gravity (EiBI) \cite{Banados:2010ix}. 	EiBI gravity belongs to the so-called Ricci Based Gravities (RBG), a family of viable gravitational extensions of GR constructed in terms of scalars out of contractions of the metric with the (symmetric part of the) Ricci tensor. RBGs are formulated \textit{\`a la Palatini}, that is, taking  metric and affine connection as independent entities. This fact allows them to yield second-order field equations that do not propagate additional degrees of freedom beyond the two polarizations of the gravitational field. This acts as a safeguard of (most of) these theories against getting into conflict with solar system tests and gravitational wave observations so far, while at the same time offering a workable framework to extract new gravitational physics in the strong-field regime. 

The relative simplicity of the stellar structure equations of EiBI gravity is a key feature that has allowed the community to scan its predictions as compared to GR expectations within different types of stars, see e.g. \cite{Pani:2011mg,Delsate:2012ky,Casanellas:2011kf,Avelino:2012ge, Banerjee:2017uwz,Wibisono:2017dkt,Danarianto:2019mxf,Rosyadi:2019hdb,Banerjee:2021auy}. In particular, on its non-relativistic regime, this theory leads to a modified Poisson equation with a single extra free parameter (for a full account of the theory and its phenomenology we refer the reader to \cite{BeltranJimenez:2017doy}). In our analysis of the pre-main sequence evolution of LMS within it, we shall mainly focus on the paths followed by the contracting star, represented by Hayashi tracks \cite{hayashi} on the Hertzsprung-Russell (HR) diagram, and the associated limiting masses by the hydrogen burning and the development of a fully convective core, respectively, at the gateway of the main sequence.  To simplify our analysis we shall disregard the deuterium burning process which happens during the pre-main sequence phase and in massive brown dwarfs, since the energy generated by this process is significantly smaller than the one of the hydrogen ignition. Another simplification of our analysis lies on the fact that in order to properly incorporate lithium burning in the low-mass stars or cooling process of a brown dwarf object, one needs to use a more realistic model of the electron degeneracy than the one used in this work, while the choice of the opacities is always subject to discussion.

This work is organized as follows: in Sec. \ref{Section2}, we introduce the main ingredients of EiBI gravity, and work out its non-relativistic equations until arriving to the generalized Lane-Emden equation for a polytropic equation of state, and set our simplified photospheric model for it and the convenctive instability criterion. In Sec. \ref{evolution}, three main aspects of  the early evolution of LMS within this theory are discussed: i) its Hayashi tracks, namely, the effective temperature vs luminosity evolution, ii) the minimum main sequence mass, namely, the minimum required mass for a star to stably burn sufficient hydrogen, and iii) the maximum fully convective mass. In all these three cases we discuss the modifications in both the positive and negative branches of the EiBI parameter (which appears unavoidably entangled with the star's central density, a common feature to RBGs), with the former being the most succulent from a physical point of view. Finally, Sec. \ref{sec:con} contains some closing thoughts.

\section{Model and basic equations }\label{Section2}

\subsection{Action and field equations of EiBI gravity} 

The action of EiBI gravity is given by
\begin{eqnarray}
\mathcal{S}_{EiBI}&=&\frac{1}{\kappa^2 \epsilon} \int d^4x \left[\sqrt{-\vert g_{\mu\nu} + \epsilon R_{\mu\nu}(\Gamma) \vert} - \lambda \sqrt{ -g}\right] \nonumber \\
&+&  \mathcal{S}_m (g_{\m \nu}, \psi_m) \label{eq:SEiBI}
\end{eqnarray}
where $\kappa^2=8\pi G/c^4$ is Newton's constant, $\e$ is the (length squared) EiBI parameter, $\lambda$ is related to the asymptotic character of the solutions (from now on we fix $\lambda=1$ to deal with asymptotically flat solutions), vertical vars denote a determinant, while we reserve the notation of $g$ for the determinant of the space-time metric $g_{\mu\nu}$, the latter being a priori independent of the affine connection  $\Gamma \equiv \Gamma_{\mu\nu}^{\lambda}$ of the (symmetric) Ricci tensor $R_{\mu \nu} (\Gamma)$. As for the matter fields, collectively labeled by $\psi_m$, they only couple to the space-time metric, but not to the independent connection $\Gamma$, which is consistent as long as no fermionic fields are present \cite{Afonso:2017bxr}. 

The field equations of action (\ref{eq:SEiBI}) are obtained as
\begin{equation} \label{eq:metric}
    \frac{\sqrt{-q}}{\sqrt{-g}} q^{\mu \nu}-g^{\m \nu}= \kappa^2 \epsilon T^{\m \nu} \ ,
\end{equation}
where the rank-two tensor defined via
\begin{equation}\label{Eq: q}
q_{\mu \nu} \equiv g_{\mu\nu} + \epsilon \, R_{\mu\nu}(\Gamma) \ ,
\end{equation}
is the metric compatible with the independent connection, while  $T_{\mu\nu} \equiv -\frac{2}{\sqrt{-g}}\frac{\delta \mathcal{L}_m}{\delta g^{\mu\nu}}$ is the stress-energy tensor of the matter. The relation \eqref{Eq: q} can be rewritten in the more seductive form
\begin{equation}\label{Eq: q Omega}
q_{\mu \nu}= g_{\mu\a} {\Omega^\a}_\n \ ,
\end{equation}
in such a way that the {\it deformation matrix} ${\Omega^\m}_\a$ is determined by the algebraic relation
\begin{equation}\label{Eq: deformation matrix}
    |\Omega|^{1/2} ({\Omega^\m}_\nu)^{-1} = \delta_{\mu}^\nu -\k^2 \e \, {T^\m}_\n \ .
\end{equation}
With the help of the definitions above, the metric field equations (\ref{eq:metric}) can be recast as
\begin{equation} \label{eq:eom}
{R^\mu}_{\nu}(q)=\frac{1}{\vert \Omega \vert^{1/2}} \left(\mathcal{L}_G \delta^\mu_\nu + \kappa^2 {T^\mu}_{\nu} \right) \ ,
\end{equation}
where the gravitational EiBI Lagrangian can also be expressed in terms of the deformation matrix as $\mathcal{L}_G=\tfrac{\vert \Omega \vert^{1/2}}{\epsilon \kappa^2}- 1$. The representation (\ref{eq:eom}) of the EiBI field equations puts forward the existence of an Einstein frame for this theory, sourced on its right-hand via additional (non-dynamical) couplings in the matter fields, a common feature of the whole RBG family \cite{Afonso:2018bpv}. In the non-relativistic limit, all these RBG theories share a common set of extra pieces to the Poisson equation with theory-dependent coefficients \cite{Olmo:2021yul}.

\subsection{Non-relativistic stellar structure equations}

Let us thus head to the non-relativistic limit of the field equations (\ref{eq:eom}) above. To this end, we set the following ansatz for the time-independent metric:
\begin{equation}
ds^2 = -(1+2\Phi)dt^2 + (1-2\psi) d\Vec{x} d\Vec{x} \ ,
\end{equation}
where $\Phi $ and $\psi$ are only functions of $\Vec{x}$. As for the stress-energy tensor, since the pressure is generally negligible for non-relativistic stars, we take a relativistic pressureless fluid, $T^{\mu \n} = \rho \,u^\m u^\n$, where $\rho$ is the energy density and $g^{\mu\nu}u_\m u_\n = -1$ a unit time-like vector. From Eq.\eqref{Eq: deformation matrix}, we can easily find the components of the corresponding deformation matrix and apply them to Eq.\eqref{Eq: q Omega} to get the metric components of $q_{\m \n}$. Finally, if one expands Eq.\eqref{Eq: q} up to linear order in $\Phi, \, \psi$, $\rho$ and their derivatives, the modified Poisson equation is found to be \cite{Pani:2011mg,Olmo:2021yul}
\begin{equation}
	\nabla^2 \Phi =  \dfrac{\kappa^2}{2}  \rho+\dfrac{\k^2 \e}{4} \nabla^2 \r \ ,
\end{equation}
where the second term corresponds to the EiBI correction. In a static, spherically symmetric space-time, the above equation can be rewritten as
\begin{equation}\label{modified Pisson Eq}
    	\dfrac{1}{r^2}\dfrac{d}{dr}\left(r^2 \dfrac{d\Phi}{dr} \right) = \dfrac{\kappa^2}{2} \rho +\dfrac{ \kappa ^2\e}{4\,r^2}\dfrac{d}{dr}\left(r^2 \dfrac{d\rho}{dr} \right) \ .
\end{equation}
Using the hydrostatic equation $\tfrac{d\Phi}{dr}=-\rho^{-1} \tfrac{dP}{dr}$ and integrating it over the radial coordinate $ r $, the above equation transforms into
\begin{eqnarray}
 \dfrac{dP}{dr} =-\dfrac{ \kappa^2 M(r) \rho}{8 \pi r^2} - \dfrac{\kappa ^2\e \,\rho}{4} \dfrac{d\rho}{dr} \ , \label{differential pressure}
\end{eqnarray}
where  the mass function $ M(r) $ is defined as
\begin{equation}\label{Eq: mass function}
    M(r)= \int^r_0 4\pi x^2 \r dx \ .
\end{equation}
The solutions of the hydrostatic equilibrium equations (\ref{differential pressure}) and (\ref{Eq: mass function}), equipped with an equation of state [given in Sec. \ref{pol}], provide the main ingredient from the gravitational sector for the internal and external features of a LMS on its early evolutionary stages. As shall be discussed later, they also contribute to the description of the boundary region between the star's interior and its photosphere, and have an impact on the photospheric quantities.

\subsection{Polytropic stars }\label{pol}

It is now time to move to the matter description of LMS. Our simplified model assumes a fully convective interior, from the center up to the photosphere. Such objects are typically well described by a polytropic equation of state which in a general case takes the form
\begin{equation}\label{eospol}
    p=K\rho^\frac{n+1}{n} \ ,
\end{equation}
where $n$ is the polytropic index, while  $K$ is the degenerate parameter which carries the information about the microscopic features of the fluid, such as e.g. electron degeneracy, Coulomb force, and ionization. Let us introduce the following dimensionless variables 
\begin{equation}\label{dimensionless variables}
\r= \r_c  \t^n ,\;\;\; P=p_c \t^{n+1},\; r=r_c \x \ ,
\end{equation}
where $\r_c$ and $p_c$ are the star's central density and pressure, respectively, while $r_c$ is defined via the expression
 \begin{equation}\label{Eq: rc}
 r_c^2= \dfrac{2 (n+1) p_c }{\k^2 \rho_c^2  } = \frac{2(n+1)K \r_c^{1/n-1}}{\k^2} \ .
 \end{equation}
These variables allow to rewrite the hydrostatic equilibrium equation \eqref{modified Pisson Eq} in a more suitable form under the modified Lane-Emden equation
\begin{equation}\label{Eq. Lane-Emden}
\dfrac{d}{d\xi}\left\lbrace  \x^2  \, \dfrac{d\t}{d\x} \left[ 1 +\alpha\t^{n-1}   \right] \right\rbrace= - \x^2 \t^n \ .
\end{equation}
In this equation the EiBI corrections are encapsulated into the single dimensionless parameter
\begin{equation}\label{Eq: def alpha}
    \alpha=\dfrac{ \e \, n}{2 \, r_c^2} \ ,
\end{equation}
which depends not only on the polytropic parameters but also on the star's central energy density $\rho_c$, as given by Eq.\eqref{Eq: rc}. This is a general feature of Palatini theories of gravity (at least for the RBG family), caused by the particular way the matter fields source the new gravitational dynamics \cite{aw_pol}, and strongly departs from what happens in other theories of gravity, including GR itself. This implies that astrophysical constraints on EiBI gravity (and on the whole RBG family) parameter requires further information on the star's central density, as we shall see later.

The generalized Lane-Emden equation \eqref{Eq. Lane-Emden} can be used to rewrite the mass function (\ref{Eq: mass function}) as well as other relevant stellar features such as radius, central density and temperature in terms of its solutions:
\begin{eqnarray}
M &=& 4 \p \r_c r_c^3 \omega_n,\label{Eq: star mass}\\
R&=& \gamma_n  \left( \dfrac{K}{G}\right) ^{\frac{n}{3-n}} M^{\frac{1-n}{3-n}} \ , \label{Eq: star radius}\\
\r_c &=&\delta_n\left( \dfrac{3M}{4\p R^3}\right) \ , \label{Eq: central dens}\\
T&=&\frac{K \m}{N_Ak_B} \r_c^{1/n}\theta \ , \label{Eq: temperature}
\end{eqnarray}
where $k_B$ is Boltzmann's constant, $N_A$ the Avogadro number and $\mu$ the mean molecular weight, while the central temperature is defined as $ T_c = \frac{K \m}{N_Ak_B} \r_c^{1/n}$. The three remaining constants, $\omega_n$, $\gamma_n$, and $\delta_n$, come from the evaluation of the corresponding solution of the generalized Lane-Emden equation (\ref{Eq. Lane-Emden}) at the star's surface,  $\xi_R$, via the expressions
\begin{eqnarray}\label{parametry}
\omega_n&=&\left[-\x^2  \, \dfrac{d\t}{d\x} \left( 1 +\alpha \t^{n-1}   \right) \right]= \left[-\x^2  \, \dfrac{d\t}{d\x} \right]_{\x=\x_R},  \\
\gamma_n&=&(4\p)^{\frac{1}{n-3}} (n+1)^{\frac{n}{3-n}} \, \omega_n^{\frac{n-1}{3-n}}  \x_R,\\
\delta_n &=& -\dfrac{\x_R}{3\left[ \dfrac{d\t}{d\x} \left( 1 +\alpha  \t^{n-1} \right) \right]}  = -\frac{\x_R}{3 \dfrac{d\t}{d\x}\Big|_{\x=\x_R}},
\end{eqnarray}
where the last equalities come from applying the boundary condition of the surface, i.e., $\theta(\x_R)=0$. 

In general, analytic solutions to the generalized Lane-Emden equation (\ref{Eq. Lane-Emden}) are not possible, so one has to resort to a numerical resolution procedure. For the sake of this work, let us take the value of the polytropic index $n=3/2$, which is the one suitable to describe the  convective interior of a LMS. In such a case, one can approximate the central behaviour of the solution of the Lane-Emden equation (\ref{Eq. Lane-Emden}) by
\begin{equation}\label{Eq: Theta}
\t(\x \approx 0)= 1- \dfrac{\x^2}{6(1+\alpha)} \sim \text{exp}\left( -\dfrac{\x^2}{6(1+\alpha)}\right),
\end{equation}
where the  initial conditions $\t (0)=1 $ and $\t' (0) = 0 $ have been applied. This result will prove its usefulness later.

\subsection{Simple photospheric model}

As already mentioned, the model described above holds up to the photosphere, which is the outer, luminous layer that delimits the star. It is formally defined as the radius for which the so-called optical depth equals the value 2/3, that is (see e.g. \cite{hansen})
 \begin{equation}\label{tau}
     \tau (r) = \int_{r_{ph}}^\infty \k_{op} \, \r \, dr=\frac{2}{3} \ ,
 \end{equation}
 where $\k_{op}$ is dubbed as the opacity, a phenomenological quantity playing a key role in the characterization of the star. Later on,  we shall use various opacity models, depending on the physical features of the material filling  the star. The photosphere is so close to the surface of the star that its radius, $r_\text{ph}$, can be well approximated by the star's radius, $R$. The photospheric temperature, often called ``effective temperature", is the one appearing in the Stefan-Boltzmann equation
\begin{equation}\label{stef}
    L=4\pi\sigma R^2 T^4_\text{eff} \ ,
\end{equation}
where $\sigma$ is the Stefan-Boltzmann constant and $L$ the luminosity. Therefore, we shall assume the star  to radiate its energy as a black body with a temperature $T_\text{eff}$.

The hydrostatic equilibrium equation \eqref{differential pressure} can be conveniently rewritten in the following way
\begin{equation}\label{Eq: hydrostatic eq}
p' = -\r\,\left( g+\dfrac{\kappa ^2\e \r'}{4} \right) \ ,
\end{equation}
where primes indicate radial derivatives, while  $ g  $ is the surface gravity defined as
\begin{equation}\label{Eq: g}
g\equiv \dfrac{\k^2 M(r)}{8\p r^2} \sim \dfrac{\k^2 M}{8\p R^2}=\text{const} \ .
\end{equation}
%write in terms of \kappa
Taking up to two derivatives in the equation above and using the definition (\ref{Eq: mass function}), one can combine the resulting expressions to find
\begin{equation}\label{Eq: rho'}
\rho'=-\dfrac{M(r)}{2 \p r^4} \ ,
\end{equation}
so that Eq.\eqref{Eq: hydrostatic eq} evaluated in the photospheric region becomes
\begin{equation}\label{Eq: hydrostatic eq 2}
p'_{ph} =-\r g \,\left(1- \frac{\e}{R^2}\right) \ ,
\end{equation}
where in this equation we have set units $\kappa^2=8\pi G$ (and assumed  $c=1$ from now on). This equation can be integrated with the help of (\ref{tau}), providing the photospheric pressure as
 \begin{equation}\label{fotopress}
     p_{ph}= \dfrac{2g\left(1- \dfrac{\e}{R^2}\right)}{3 \k_{op}} \ .
 \end{equation}
It is clear now that the most relevant element of the photosphere's modelling is its opacity. Depending on the physical conditions, mainly contained within the pressure and temperature regimes of the considered stages of the stellar evolution, we shall use different analytical expressions, which approximately reflect how opaque matter is to the electromagnetic radiation.

\subsection{Convective instability - modified Schwarzschild criterion}\label{Sec: Swch}

Another crucial information in the description of stellar interiors is how the energy is transported through different regions of a star. Since our LMS is modelled by a fully convective sphere enveloped by a radiative photosphere, one needs a formal criterion encapsulating the physical conditions responsible for any of those energy transports. This is given by the so-called Schwarzschild criterion, turning out to be dependent on the underlying theory of gravity, as shown in \cite{Wojnar:2020ckw}. Therefore, the heat is transported via radiative processes when the temperature gradient is smaller than the adiabatic one
\begin{equation}\label{Schw criterion}
\nabla_{rad}< \nabla_{ad} \ .
\end{equation}
For the rest of our setup we shall model the photosphere's matter as an ideal, monoatomic gas, for which it can be shown that the adiabatic gradient has a constant value, $ \nabla_{ad} =0.4 $ \cite{hansen}. On the other hand, the radiative gradient is defined as 
\begin{equation}\label{rad}
\nabla_{rad}:=\left(\dfrac{d \ln T}{d \ln P} \right) _{rad}.
\end{equation}
To find its form and dependence on EiBI gravity parameter, we need to analyze the radiative heat transport equation, which is given by
\begin{equation}\label{Eq: radiative heat transoprt}
\dfrac{\partial T}{\partial m}=-\dfrac{3}{64\pi^2 a} \dfrac{\k_{rc} l}{r^4 T^3} \ ,
\end{equation}
where $ \k_{rc} $ is the radiative or/and conductive opacity, $ l $ the local luminosity, while $ a=7.57 \times 10^{-15}\frac{erg}{cm^3 K^4} $ represents the radiation density constant. Combining the above expression with the modified hydrostatic equilibrium equation \eqref{differential pressure} differentiated with respect to the mass, one gets
\begin{equation}
\dfrac{\partial T}{\partial P}= \dfrac{3 \k_{rc}\, l}{16\pi r^2 a T^3} \left( \dfrac{Gm}{r^2}+ \dfrac{\k^2 \e \r'}{4 } \right)^{-1} \ .
\end{equation}  
%write in terms of \kappa
Using this result into the definition (\ref{rad}) provides the radiative temperature gradient for EiBI gravity as
\begin{equation}\label{Eq: radiative temp gradient}
\nabla_{rad}= \dfrac{3 \k_{rc}\, l \,p}{16\pi r^2 a T^3} \left( 1- \dfrac{ \e }{r^2 } \right)^{-1} \ ,
\end{equation}
and therefore, depending on the value of the EiBI parameter $\e$, this modification has a (des-)stabilizing effect, altering a radiative region development.

\section{Early evolution of low-mass stars}\label{evolution}

With the formalism developed in the previous section, we are ready to study the early evolution of a LMS within EiBI gravity. When the baby proto-star approaches the main sequence, it is a luminous but otherwise cold (sub-) stellar object. Similarly as for the later phases, such an object can be accommodated on the HR diagram; thus, it can be found on the right-hand side part of the evolutionary diagram, above the main sequence band. The evolutionary path that it follows is called a Hayashi track \cite{hayashi}, described by a relation between the effective temperature, luminosity, mass, and metallicity, where the last one is responsible for the shape of the curve. However, since we are dealing with a toy-model description to understand the new features brought by the gravitational corrections of the EiBI gravity, that aspect will not be apparent in our subsequent analysis.

A stellar object will leave its Hayashi track when any of the following processes happens:
\begin{itemize}
    \item { \it Radiative core development:} Since the luminosity decreases as the baby star follows the Hayashi track down but the effective temperature remains almost constant, this means that, from the Stefan-Boltzman law, the star is contracting. Therefore, it may happen that the star's interior becomes radiative, as a consequence of increasing its interior temperature. In such a situation, it will reach a minimum and follow an almost horizontal line before getting to the main sequence, moving to higher effective temperatures. This stage of the early evolution is called a Henyey track \cite{henyey,hen2,hen3}, and it will not be studied here; however, in Sec. \ref{sec:MFCM} we shall discuss in detail the onset of the radiative core development as a boundary condition of the fully convective star on the main sequence.
    \item { \it Hydrogen ignition:} When the central temperature and pressure increase, the conditions present in the stellar core can become sufficient to ignite hydrogen and stop further gravitational contraction. If the process is stable, in other words, when the energy radiated away through the photosphere is balanced by energy produced by the hydrogen burning in the core, the star has evolved to the next stage of the stellar evolution, that is, the main sequence phase, which we analyze in Sec. \ref{sec:MMSM}.  
    \item {\it Contraction stops at the
    onset of electronic degeneracy:} This process will happen when none of the above ones takes place - that is, the interior of such an object is too cold to start hydrogen burning. Apart from the light elements burnt in the initial phase, those objects do not possess any source of energy production in their cores and, therefore, they will cool down with time when electron degeneracy pressure balances the gravitational contraction \cite{Burrows:1992fg}. Such objects are called brown dwarfs, and will be discussed somewhere else in detail.
\end{itemize}

\subsection{Hayashi tracks} \label{sec:Hay}

In what follows, we will now focus on a simple description of the Hayashi tracks in EiBI gravity. As mentioned before, the objects following this stage of the evolution are fully convective, and we shall also assume that their interiors are made of a fully ionized monatomic gas with temperature $T$ and mean molecular weight $\mu$. In such a situation, the equation of state can still be formally recast as polytropic (\ref{eospol}):
\begin{equation}\label{Eq: polytrop}
p= \tilde{K} T^{1+n} \ ,
\end{equation}
where the polytropic constant $\tilde K$ is related to the degenerate one by
\begin{equation}\label{Eq: polytropic constant}
 \tilde{K} = \left( \dfrac{N_A k_B}{\m}\right) ^{-(n+1)} K^{-n} \ .
\end{equation}
Note that in the relation (\ref{Eq: polytrop}) we have used the ideal gas law given by
\begin{equation}
    \rho = \frac{\mu p}{N_A k_B T} \ .
\end{equation}
It is worth stressing  that $K$ depends on the theory of gravity, since it can be expressed with respect to the solution of the modified Lane-Emden equation (\ref{Eq. Lane-Emden}) via
\begin{equation}\label{ka}
 K=\left[\frac{4\pi}{\xi_R^{n+1}(-\theta'_n(\xi_R))^{n-1}}\right]^\frac{1}{n}\frac{G}{n+1}M^{1-\frac{1}{n}}R^{\frac{3}{n}-1} \ .
\end{equation}
Note also that the equation of state (\ref{Eq: polytrop}) is valid up to the photosphere, since above the interior-photosphere boundary the energy transport is ruled by radiative processes instead. In such a region, we shall use a simplified relation for the absorption law, given by the Kramer formula:
\begin{equation}\label{Eq: opacity}
 \kappa_\text{abs}=\kappa_0p^i T^j \ .
\end{equation}
For cold stars, whose effective temperatures lie inside the range $3000\lesssim T\lesssim6000$K, the surface layer
is dominated by H$^{-}$ opacity \cite{hansen}. Considering the hydrogen mass fraction as $X\approx0.7$, the opacity is given by
\begin{equation}\label{hydro}
 \kappa_{H^-}=\kappa_0 \rho^\frac{1}{2}\,T^9\,\,\text{cm}^2\text{g}^{-1} \ ,
\end{equation}
with $\kappa_0\approx2.5\times10^{-31} \left(\frac{Z}{0.02}\right)$, where the metal mass fraction $Z$ (or metallicity) is an important element in the stellar modelling. Its value is typically taken within the range $0.001\lesssim Z\lesssim0.03$ \cite{cantiello}; as an example, the solar metallicity is $Z=0.02$. Since that region can be also modelled as an ideal gas,  the opacity (\ref{hydro}) can be expressed as
\begin{equation}\label{hydro2}
 \kappa_{H^-}=\kappa_g p^\frac{1}{2}\, T^{8.5}\,\,\text{cm}^2\text{g}^{-1} \ ,
\end{equation}
where we have redefined $\kappa_g=\kappa_0\left(\tfrac{\mu}{N_Ak_B}\right)^{\frac{1}{2}}\approx1.371\times10^{-33}Z\mu^\frac{1}{2}$. Particularizing the relation  (\ref{tau}) to ${H^-}$ opacity, one finds that the photospheric pressure is
\begin{equation}\label{Eq: pre photosphertic p}
     p_{ph}= \dfrac{2g\left(1- \dfrac{\e}{R^2}\right)}{3 \k_{H^-}} \ .
 \end{equation}
Applying the solution of the modified Lane-Emden (\ref{Eq. Lane-Emden}) for $n=3/2$ to the above expression, the Stefan-Boltzmann law (\ref{stef}) and the opacity expression (\ref{hydro2}), the photospheric pressure above takes the form
\begin{equation}\label{Eq: photospheric p}
 	p_{ph}=8.11279\times 10^{14} \left[ \frac{M \,\beta }{\sqrt{\mu } L T^{4.5} Z}\right] ^{2/3},
\end{equation}
where we have redefined the brackets appearing in Eq.\eqref{Eq: pre photosphertic p} as
\begin{equation} \label{eq:beta}
\beta=1-\frac{2 \alpha }{ 3^{5/3} \delta ^{2/3} \omega^{2/3}}.
\end{equation}
For simplicity, we have removed the sub-indices $n$ appearing in the relations (\ref{parametry}) and from now on we will understand them as their values for $n=3/2$.

\begin{figure}[t!]
	\centering
	\includegraphics[width=1\linewidth]{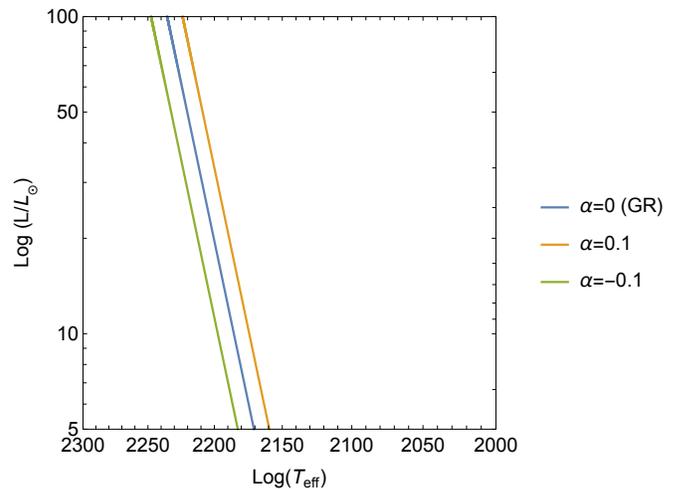}
	\caption{The piece of the HR diagram representing shifted Hayashi tracks by the modifications introduced by EiBI gravity model in logarithmic scale. The curves are given by the equation (\ref{hay}) taking $M/M_{\odot}=1/2$, for some chosen values of the parameter $\alpha$ defined in Eq.(\ref{Eq: def alpha}) as compared to the GR/Newtonian curve, $\alpha=0$.}
	\label{fig:hayashi}
\end{figure}

The above photospheric pressure must be matched to the pressure of the ideal gas given by the relation (\ref{Eq: polytrop}) evaluated at the photosphere. The latter yields, after using the Stefan-Boltzmann law (\ref{stef}), the effective temperature under the form
\begin{equation}\label{Teff}
	T_{eff}= 9.1960 \times 10^{-6} \left( \dfrac{\mu ^5 L^{3/2} M p_{ph}^2}{-\theta ' \xi _r^5}\right) ^{1/11} \ ,
\end{equation}
which after using the derived photospheric pressure \eqref{Eq: photospheric p} provides the result
\begin{equation}\label{hay}
    T_{ph}=2482.10 \mu ^\frac{13}{51} \left(\frac{L}{L_{\odot}}\right)^\frac{1}{102} \left(\frac{M}{M_{\odot}}\right)^\frac{7}{51} \frac{ \beta^\frac{4}{51}}{Z^\frac{4}{51}(\sqrt{-\theta '} \xi _R^{5})^\frac{1}{17}} \ ,
\end{equation}
where we have re-scaled both mass and luminosity to their solar values, $\{M_{\odot},L_{\odot}\}$. Let us notice that the numerical value in the above expression is too low; it should be almost a twofold larger. The reason of this reduced value lies in the simplifications we have made, mainly related to the photospheric modelling. Notwithstanding, this analytical formula allows us to track down the modifications introduced by EiBI gravity to the early stage of the stellar evolution. Therefore, for a given star with mass $M$, uniform mean molecular weight $\mu$, and metallicity $Z$, the above expression gives the corresponding  Hayashi track. These almost vertical lines are evolutionary tracks of infant stars with masses supposedly below  $\sim 0.5M_\odot$, though such a limiting mass also depends on the theory of gravity. Its shape and position on the HR diagram depend not only on the metallicity, but also on the theory of gravity, which in the present case is encapsulated in the parameter $\beta$ appearing in Eq.(\ref{eq:beta}) but also through the solutions of the extended Lane-Emden equation (\ref{Eq. Lane-Emden}). This gravity model-dependence could actually allow us to constraint the theory's parameter when a more realistic atmosphere analysis is performed, since none such star is placed in the so-called Hayashi forbidden zone, being found in the low temperatures. As presented in Fig. \ref{fig:hayashi}, EiBI gravity shifts the tracks in opposite ways depending on the sign of the theory's parameter $\epsilon$, either in the direction of the forbidden zone (for $\epsilon>0$), which lies in the region of lower temperatures, or against it (for $\epsilon<0$). 

\subsection{Minimum main sequence mass}\label{sec:MMSM}

\begin{figure}[t!]
	\centering
	\includegraphics[width=1\linewidth]{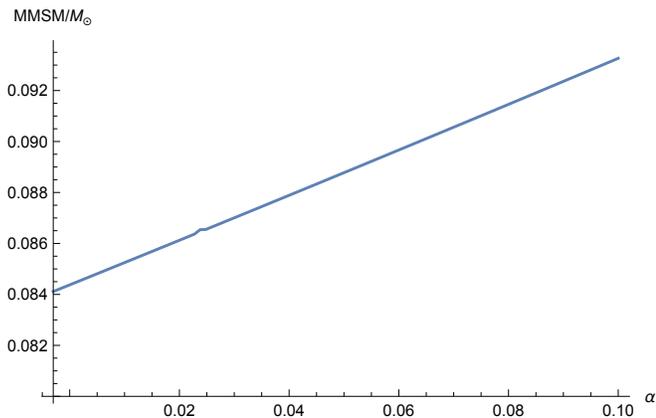}
	\caption{The evolution of the MMSM (in units of solar masses) with the parameter $\alpha$. The range in this plot goes from $\alpha \in (-0.003,0.10)$, with the lower bound given by a well-defined solution of the extended Lane-Emden equation (\ref{Eq. Lane-Emden}).}
	\label{fig:MMSM}
\end{figure}

Using the ingredients introduced in the previous section, we are now capable to  compute the minimum main sequence mass (MMSM). This is the minimal mass required by a star to ignite sufficiently stable thermonuclear reactions in its interior to compensate photospheric energy losses. Even though the central temperature can be sufficient to start the p-p chain, it is not necessarily enough to complete it. The thermonuclear rates depend mainly on the temperature and density, in such a way that the energy generation rate can be well approximated by power laws of the form (see \cite{Burrows:1992fg} for details)
\begin{equation}\label{Eq: energy generation rate}
\dot{\epsilon}_{pp} = \dot{\epsilon}_c \left( \frac{T}{T_c}\right) ^s\left( \frac{\r}{\r_c}\right) ^{u-1} \ ,
\end{equation}
where the two exponents can be phenomenologically fitted as $ s\approx 6.31 $ and $ u\approx2.28 $ at the transition mass of the core, while the function 
\begin{equation}
    \dot{\e}_c= \dot{\e}_0 \,T_c^s \, \r_c^{u-1} \, \text{ergs g}^{-1}\text{s}^{-1} \ ,
\end{equation}
%la\e_0 esta mal
with $\dot{\e}_0 \approx 3.4 \times 10 ^{-9} $ in suitable units. The corresponding luminosity of the hydrogen burning is found as
\begin{equation}\label{Eq: Luminosity}
L_{pp}=\int \dot{\epsilon}_{pp}\, dM=  4\p \dot{\epsilon}_c r_c^3 \r_c  \int_{0}^{\x_R} \t^{n\left(u+\frac{2}{3}s\right)}\, \x^2 d\x  \ ,
\end{equation}
where we have used the fact that $(T/T_c)=(\rho/ \rho_c)^{2/3}$ along the adiabatic core. The last integral can be easily computed by using the approximation \eqref{Eq: Theta} and the definition (\ref{Eq: star mass}), which yields the result
\begin{equation}\label{Eq: Integral Lpp}
L_{pp}=\dfrac{6 \sqrt{3\pi (\alpha +1)^3 }}{\omega_{3/2} (2 s+3 u)^{3/2}}\dot{\epsilon}_c M \ .
\end{equation}
In this approximation, we have taken into account the fact that most of the hydrogen burning will be produced in a region near to the core of the star. Now, considering  that the fraction of hydrogen in a high-mass brown dwarf is of $75\%$,  and that  the number of barions per electron can be approximated to $\mu_e = 1.143$, besides setting the following degenerate polytropic constant
\begin{equation}
    K= \dfrac{ (3\pi^2)^{2/3} \hbar}{5 m_e m_H^{5/3} \m_e^{5/3}} \left( 1+ \frac{\a_d}{\eta} \right),
\end{equation}
where $\hbar$ is the reduced Plank constant, $m_e$ is the electron mass and $m_H$ is the proton mass. Then, the luminosity (\ref{Eq: Integral Lpp}) can be recast as
\begin{equation} \label{eq:Lpp}
	L_{pp} =1.54 \times 10^7  L_ \odot \dfrac{\delta^{5.49} \,(1+ \alpha )^{3/2}}{ \gamma ^{16.46} \omega} M_{-1}^{11.97}  \,\dfrac{\eta^{10.15}}{ \left(\alpha_d +\eta \right)^{16.46}} \ ,
\end{equation}
where we have defined here, by convenience, $M_{-1}=M/(0.1M_{\odot})$. In order to find the MMSM we need to equal this hydrogen burning luminosity to the one of the photosphere. For the purpose of computing the latter, we take Eq.\eqref{Eq: pre photosphertic p} and assume again that the components of the stellar atmosphere behave as ideal gas, that is
\begin{equation}\label{Eq: ideal gas}
\dfrac{ \rho _{ph} k_B T_{ph}}{\mu  m_H}=\frac{2 g \,\left(1- \dfrac{\e}{R^2}\right)}{3 \kappa_{op}} \ .
\end{equation}
This equation will allow us to get a relation between $\r_{ph}$ and $M$, but before going that way, let us first rewrite the surface gravity $ g $ defined in Eq.\eqref{Eq: g} as
\begin{equation}\label{Eq: g final}
g=\frac{G^3 M^{5/3}}{\gamma^2 K^2} \ ,
\end{equation}
and consider that the photospheric temperature can be found from the matching of the specific entropies of the gas/metallic phases there, which yields \cite{Burrows:1992fg}
\begin{equation}\label{Eq: photospheric temperature}
T_{ph}=\frac{1.8 \times 10^6 \rho _{ph}^{0.42}}{\eta ^{1.545}} \ .
\end{equation}
Replacing the above two equations into \eqref{Eq: ideal gas}, we find
\begin{equation}\label{Eq: photospheric density}
\rho _{ph}=2.957 \times 10^{-5} \dfrac{\eta ^{1.09} G^{2.11} M^{1.17} \left(\mu  m_H \b \right)^{0.70}}{\left(\gamma K\right)^{1.41}(k_B \kappa _{op})^{0.70}} \ .
\end{equation}
Inserting this result back to Eq.\eqref{Eq: photospheric temperature}, the photospheric temperature becomes
\begin{equation}\label{Eq: final photospheric temp}
T_{ph}=2.254 \times 10^4 \frac{G^{0.89} M^{0.49} \left(\mu  m_H \beta\right)^{0.30}}{\eta ^{1.09} \left(\gamma K\right)^{0.59}(k_B \kappa _R)^{0.30}} \ .
\end{equation}
Therefore, the photospheric luminosity, given by $L_{ph}= 4\pi R^2 \sigma T_{ph}^4$, can be expressed in terms of the star mass, $M$, as follows
\begin{equation}\label{luminisity}
L_{ph}=	0.534 L_\odot \dfrac{ M_{-1}^{1.31}\beta^{1.18}}{\eta ^{3.99} \gamma^{0.37} (\alpha_d +\eta )^{0.37} \kappa_{-2}^{1.18}} \ ,
\end{equation}
where we have defined the quantity $\kappa_{-2}=\kappa_R/(10^{-2} \text{cm}^2\text{g}^{-1})$. Finally, equalling the hydrogen burning luminosity (\ref{eq:Lpp}) with the photospheric one (\ref{luminisity}), we find the following expression 
\begin{equation} \label{eq:MMSM}
   M_{-1}^{\text{MMSM}}=0.227\dfrac{ \gamma^{1.51} \omega^{0.09} (\alpha_d +\eta )^{1.51} \beta^{0.11}}{(\alpha +1)^{0.14}\delta^{0.51} \eta ^{1.33}\kappa_{-2}^{0.11}},
\end{equation}
where the EiBI dependences enter in this expression both via the coefficient $\alpha$ in Eqs.(\ref{Eq: def alpha}) and Eq.(\ref{eq:beta}) and via the parameters $\{\omega,\gamma,\delta\}$ obtained from the resolution of the extended Lane-Emden equation (\ref{Eq. Lane-Emden}). This is the MMSM for EiBI gravity under the assumptions and simplifications above. In order to compute it for different values of the EiBI parameter, the main obstacle here is the fact that $\alpha$ depends on the central density of the star, as can be seen from Eqs.\eqref{Eq: rc} and \eqref{Eq: def alpha}. Therefore, for the sake of our calculations we shall take the maximal 
value for the central density of $\rho_c \sim 10^3 \text{g/cm}^3$ \cite{Burrows:1992fg}, which allow us to compute the MMSM and thus set bounds on the size of  $\alpha$ as coming from observational constraints. In Table \ref{Taula1} we actually compute the set of $\{\gamma,\omega,\delta\}$ values for several choices of the parameter $\alpha$ in order to find the corresponding MMSM. For $\alpha=0$ (GR case) we get $M^{\text{MMSM}} \approx 0.084M_{\odot}$, which is somewhat halfway between other analytical calculations \cite{Burrows:1992fg} and the results of numerical simulations \cite{Kumar}. For non-vanishing values of $\alpha$, this table shows that for positive (negative) $\alpha$ the MMSM is larger (smaller). Thus, the positive branch of $\alpha$ is the most interesting one for our purposes, since it allow us to constrain its size via comparison with the observations of the less massive main-sequence stars ever observed, which corresponds to the $0.0930 \pm  0.0008M_{\odot}$ of the M-dwarf star G1 866C \cite{Segransan}. This way, in Fig. \ref{fig:MMSM} we numerically depict the evolution of the MMSM with $\alpha>0$. Our results within our simplified model points that near values of the parameter $\alpha \gtrsim 0.1$ the model is likely to run into conflict with observations, therefore setting a bound to the combination of the EiBI parameter and the star's central density, the latter to be estimated by other means. On the other hand, in the negative branch we run into a problem related to the fact that when the parameter reaches $\alpha \lesssim -0.003$ there are non-physical solutions, since below that value the sign would change in the bracket of the extended Lane-Emden equation (\ref{Eq. Lane-Emden}). Let us however note that, due to the same reasons stated above, this feature does depend on the star's central density, hence for a less dense or a denser core we would deal with different singular values of the parameter and, therefore, we do not extract any conclusion on the limit of validity of this branch within the formalism presented here.

\begin{table}[t!]
\begin{center}
	\begin{tabular}{|c|c|c|c|c|}
	\hline
	$ \alpha $ & $ \gamma_{3/2} $ & $  \omega_{3/2} $ & $ \delta_{3/2} $&  $ \text{MMSM}/M_\odot $\\
	\hline
	0.100 & 2.49 & 3.01 & 5.74 & 0.0933 \\
	0.010 & 2.37 & 2.74 & 5.96 & 0.0852 \\
	0.001 & 2.36 & 2.72 & 5.99 & 0.0845 \\
	0 (GR) & 2.36 & 2.71 & 5.99 & 0.0844 \\
	-0.001 & 2.36 & 2.71 & 5.99 & 0.0843 \\
    -0.003 & 2.35 & 2.69 & 6.03 & 0.0837 \\
\hline
\end{tabular}
\end{center}
\caption{The MMSM (in units of solar masses) computed with Eq.(\ref{eq:MMSM}) for several values of the parameter $\alpha$ defined in (\ref{Eq: def alpha}), including the intermediate values of the parameters $\{\gamma,\omega,\delta\}$ obtained from the resolution of the extended Lane-Emden equation (\ref{Eq. Lane-Emden}).}
\label{Taula1}
\end{table}

\subsection{Fully convective stars on the main sequence and radiative core development}\label{sec:MFCM}

 We will now focus on the final parts of the Hayashi tracks. Recalling that during this evolutionary phase the proto-star is fully convective, it might happen that the inner temperature increases enough to satisfy the conditions for radiative core development and, therefore, the object can have much more complex structure than the one we consider. Because of that, as discussed in Sec.\ref{evolution}, the star can either enter the Henyey evolutionary phase, represented by the almost horizontal to the main sequence lines, or it can stop contracting on the onset of the radiative core development as the nuclear processes start balancing the gravitational attraction. In the last situation, which is our concern now, the star begins its main sequence evolutionary phase, being however still fully convective. Using the previous results found in this paper, we can obtain the Maximum Fully Convective Mass (MFCM) of a star on the main sequence.
 
 In order to determine when the radiative processes take over in the core, one needs to analyze the Schwarzschild criterion, which tells us that the radiative processes start when $ \nabla_{ad}= \nabla_{rad} $. In our simplified modelling above we assumed that the star is made of an ideal, monoatomic gas, providing that $ \nabla_{ad}=0.4 $, while the radiative temperature gradient was already derived in Eq.\eqref{Eq: radiative temp gradient}. Applying the homology law, together with Eqs.\eqref{Eq: polytrop} and \eqref{Eq: polytropic constant}, we can express the latter as
 \begin{equation}
\nabla_{rad}= 5.21177\times 10^{69} \frac{ L \xi ^5(- \theta') \kappa _o}{\mu ^5 M^2 R^3 T^{3.5} \beta} \ ,
 \end{equation}
where $L$ is the local luminosity (here evaluated at the core). Substituting the central temperature from \eqref{Eq: temperature} and subsequently the Stefan-Boltzmann law, the above expression yields
\begin{equation}\label{Eq: gradient}
\nabla_{rad}=8.99 \times 10^{-13}  \left(\frac{L}{L_\odot}\right)^{1.25}  \frac{ \xi^{10.83} \left(-\theta '\right)^{2.17} \kappa_0 }{\delta ^{2.33} \mu ^{8.5} M_{-1}^{5.5} T} \ .
\end{equation}
Equaling this result with $\nabla_{ad}= 0.4$ one finds the 
maximum luminosity for a fully convective star on the onset of the radiative core development
\begin{equation}\label{Eq: fully conv lum}
L=2.0827\times 10^9 L_\odot \dfrac{ \beta ^{0.8} \delta ^{1.87} \mu ^{6.8} T^{0.8}}{\xi ^{8.67} \left(-\theta '\right)^{1.73} \kappa _0^{0.8}} M_{-1}^{4.4} \ .
\end{equation}
Now, by equaling this luminosity to the one of the hydrogen burning given by Eq.\eqref{luminisity} yields the MFCM
\begin{equation}\label{Eq: MFCM}
M_{-1}=1.91\dfrac{  \beta ^{0.11} \gamma ^{2.17} \mu ^{0.90} T^{0.11} \omega ^{0.13} (\a_d+\eta )^{2.17}}{\left(\alpha +1\right)^{0.20} \delta ^{0.48} \eta ^{1.34} \xi ^{1.14} \left(-\theta '\right)^{0.23} \kappa _0^{0.11}} \ ,
\end{equation}
where similar comments as on the sources of EiBI corrections of the MMSM above apply here. To make quantitative estimates of this mass, let us first assume the usual values for LMS as $ \a_d =4.82 $, $ \eta=9.4 $, $ \m=0.618 $ and $ T_\text{eff}= 4000 $K. In addition, we have to set the opacity, keeping the Kramers' form written in Eq.\eqref{Eq: opacity} with $i=1$ and $j=-4.5$; there are the total bound-free and the free-free opacities (see e.g. \cite{hansen} for details)
\begin{eqnarray}
\k_0^{bf} &\approx& 4\times 10^{25} \m \dfrac{Z(1+X)}{N_A k_B} \, \text{cm}^2 \text{g}^{-1}, \label{bound-free}\\
\k_0^{ff} &\approx& 4\times 10^{22} \m \dfrac{(X+Y)(1+X)}{N_A k_B}  \, \text{cm}^2 \text{g}^{-1}\label{free-free}.
\end{eqnarray}

\begin{table}[t!]
\begin{center}
	\begin{tabular}{|c|c|c|}
	\hline
	$ \alpha $ &  $ M_{bf}/M_\odot $&$ M_{ff}/M_\odot$\\
	\hline
 0.100 & 0.108 &0.225\\
 0.010 & 0.0994 &0.207\\
 0.001 & 0.0985 & 0.205\\
 0.000 & 0.0984 & 0.205\\
 -0.001& 0.0983 & 0.204\\
-0.003 & 0.0976 & 0.203 \\
\hline
\end{tabular}
\end{center}
\caption{Numerical values for the MFCM (in solar mass units), using the total bound-free and the free-free opacities, defined in Eq. \eqref{bound-free} and \eqref{free-free}, respectively, for different values of the composite EiBI parameter $\alpha$ appearing in Eq.(\ref{Eq: def alpha}).}
\label{tab22}
\end{table}

\begin{figure}[t!]
	\centering
	\includegraphics[width=1\linewidth]{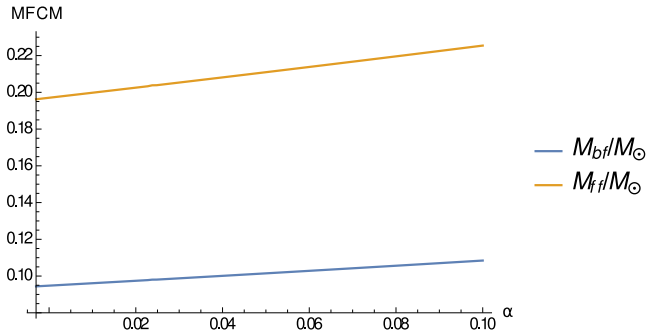}
	\caption{The dependence of the (normalized) MFCM, for both opacity models \eqref{bound-free} and \eqref{free-free}, on the parameter $\alpha \in (-0.003,0.10)$.}
	\label{fig:MFCSM}
\end{figure}

Once everything is settled, in Table \ref{tab22}  we calculate the MFCM for several values of the parameter $\alpha$ for both opacity models. Similarly as with the MMSM above, the MFCM increases (decreases) with positive (negative) gravitational parameter (note that the parameters $\{\omega,\gamma,\delta\}$ are those appearing in Table \ref{Taula1}). In addition, in Fig. \ref{fig:MFCSM} we depict the evolution with $\a$ of the two MFCM masses, corresponding to each opacity. In both table and plot it is clearly seen that the choice of the opacity model significantly affects (roughly a factor two) the value of the MFCM. As for the negative branch, we find the same feature as with the MMSM, namely, the fact that for $\a \lesssim -0.003$ the extended Lane-Emden equation \eqref{Eq. Lane-Emden} fails to provide a non-singular solution and, as such, those values are disregarded in our analysis of the MFCM.

\section{Discussion and conclusion} \label{sec:con}

In this work we have discussed several aspects of the early evolutionary phases of low-mass stars within an extension of GR dubbed as Eddington-inspired Born-Infeld gravity. Such an extension is governed by a single parameter which manifests, in the stellar structure equations of non-relativistic stars, as an extra piece to the Poisson (Lane-Emden) equation when a polytropic equation of state is considered. Supplemented with a simplified photospheric model, and equipped with a criterion for conventive instability, we investigated three features of such an early evolution of LMS.

The first feature deals with the effective temperature-luminosity relations in the evolutionary path of a proto-star, the so-called Hayashi tracks. We have shown that positive (negative) values of the EiBI parameter shift the corresponding Hayashi track in the sense of larger (smaller) effective temperature for a fixed luminosity. The second feature is the minimum required mass for a star to stably burn enough hydrogen to compensate photospheric losses, allowing it to belong to the main sequence. In this case, positive (negative) values of the EiBI parameter yield larger (smaller) minimum main-sequence masses, the former allowing to place constraints on the theory's parameter via comparison with the lowest-mass main-sequence stars every observed. This poses a difficulty for this theory, since such constraints act upon a combination of the EiBI parameter {\it and} the star's central density. This dependence of the stellar features not only on global quantities (such as the total mass) but also on local ones is a common feature of the RBG family, therefore forcing us to live with it. The third feature deals with the development of a radiative core at the end of the Hayashi track, entering the main-sequence phase while still being fully convective. We found the maximum value of the mass for this to happen, again observing an increase (decrease) of this mass with positive (negative) EiBI parameter. Note, however, that for all these three features the main astrophysical ingredient determining their absolute values is the opacity, whose modelling is always a delicate issue. Its influence is obvious in the last feature (the MFCM), where two different models of opacities (bound-free and free-free ones) result in up to a factor two in the absolute value of this quantity.

The results found in this work highlight the viability of using metric-affine gravities of the RBG type to study modifications to the stellar model predictions of GR, in particular, within the non-relativistic regime. This is so because in such a regime, RBG modifications to the usual Poisson equation typically occur via a single additional parameter \cite{Olmo:2021yul}, allowing to study the phenomenology of several types of stars, particularly low-mass stars, without ruining the consistence of the theory with weak-field limit observations. As mentioned above, the main bottleneck in order to place observational constraints upon any such theories is the determination of the central density, which up to now we have been only able to fix by taking its assumed values within GR, though more reliable theoretical procedures to deal with this issue are being investigated. We are also working in other aspects of non-relativistic objects and low-mass stars in different RBGs, and we hope to report on this soon.

\vspace{5mm}
\section*{Acknowledgements}

MG is funded by the predoctoral contract 2018-T1/TIC-10431 and acknowledges further support by the European Regional Development Fund under the Dora Plus scholarship grants. DRG is funded by the Atracci\'on de Talento Investigador programme of the Comunidad de Madrid (Spain) No. 2018-T1/TIC-10431, and acknowledges further support from the Ministerio de Ciencia, Innovaci\'on y Universidades (Spain) project No. PID2019-108485GB-I00/AEI/10.13039/501100011033, and the
FCT projects No. PTDC/FIS-PAR/31938/2017 and PTDC/FIS-OUT/29048/2017. AW is supported by the EU through the European Regional Development Fund CoE program TK133 ``The Dark Side of the Universe".   This work is also supported by the Spanish projects Nos.  FIS2017-84440-C2-1-P and PID2020-117301GA-I00, the project PROMETEO/2020/079 (Generalitat Valenciana), and the Edital 006/2018 PRONEX (FAPESQPB/CNPQ, Brazil, Grant 0015/2019). MG thanks the Laboratory of Theoretical Physics at the University of Tartu, and AW the Departament of Theoretical Physics at the Complutense University of Madrid, for their kind hospitality while doing this work.

\end{document}